\documentclass[twocolumn,aps,prl,superscriptaddress,amsfonts]{revtex4-1}

\usepackage{amsmath}
\usepackage{graphicx}
\usepackage{bm}
\usepackage{array}
\usepackage{dcolumn}
\usepackage{hepparticles}
\usepackage{heppennames}

\begin{document}

\DeclareRobustCommand{\Hbar}{\HepAntiParticle{H}{}{}\xspace}
\DeclareRobustCommand{\H}{\HepParticle{H}{}{}\xspace} \newcommand{\pbar}{\APproton}

\DeclareRobustCommand{\pbar}{\HepAntiParticle{p}{}{}\xspace}
\DeclareRobustCommand{\p}{\HepParticle{p}{}{}\xspace}

\DeclareRobustCommand{\pos}{\HepParticle{e}{}{+}\xspace}
\DeclareRobustCommand{\e}{\HepParticle{e}{}{}\xspace}

\newcommand{\zhat}{\bf \hat{z}}
\newcommand{\yhat}{\bf \hat{y}}
\newcommand{\xhat}{\bf \hat{x}}

\newcommand{\MagneticMomentHistoryFigure}{
\begin{figure}[htbp!]
\includegraphics*[width=\columnwidth]{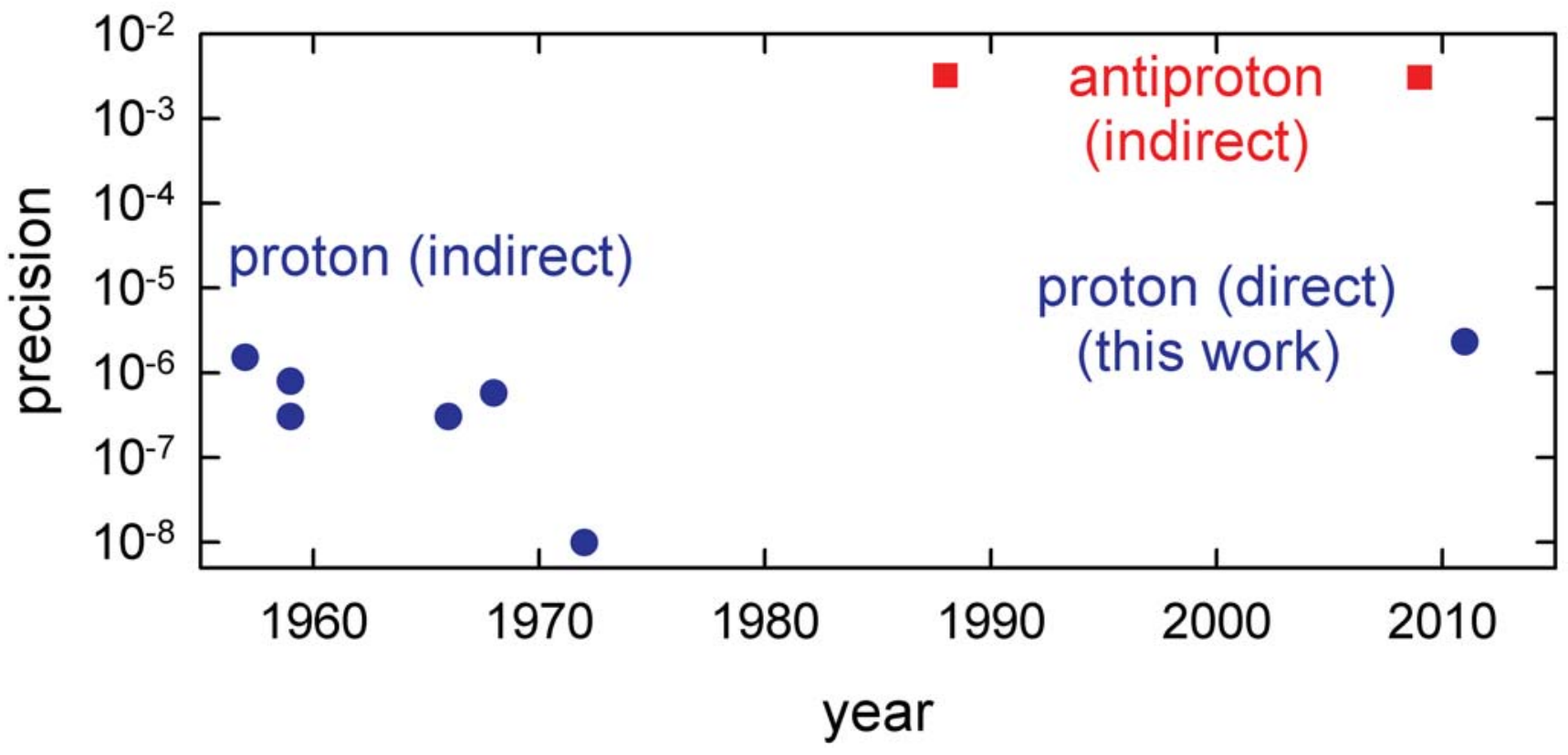}
\caption{Precision in $g_p/2$ and $g_{\pbar}/2$ for \p and \pbar \cite{MITProtonMoment,CODATA2006,1988PbarMoment,2009PbarMoment}.}
\label{fig:MagneticMomentHistory}
\end{figure}
}

\newcommand{\PbarMeasurementTrapFigure}{
\begin{figure}[htbp!]
\centering
\includegraphics*[width=\columnwidth]{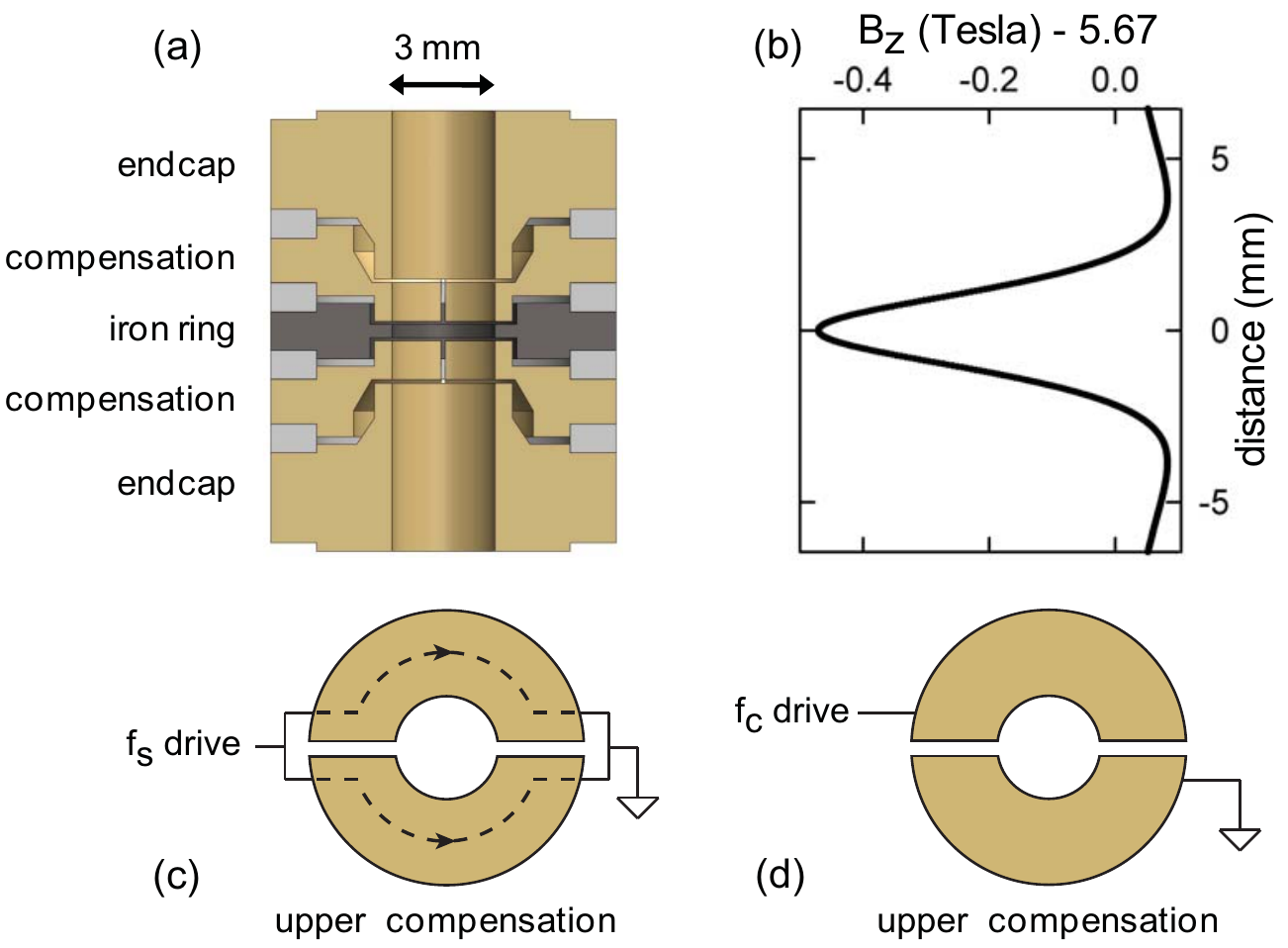}
\caption{(a) Analysis trap electrodes (cutaway side view) are copper with an iron ring. (b) The iron ring significantly alters B on axis. (c) Oscillating current paths (top view) for the spin flip drive.  (d) An oscillating electric field (top view) drives proton cyclotron motion.}
\label{fig:PbarMeasurementTrap}
\end{figure}
}

\newcommand{\HistogramRadiusAllanFigure}{
\begin{figure}[htbp!]
\includegraphics*[width=1.0\columnwidth]{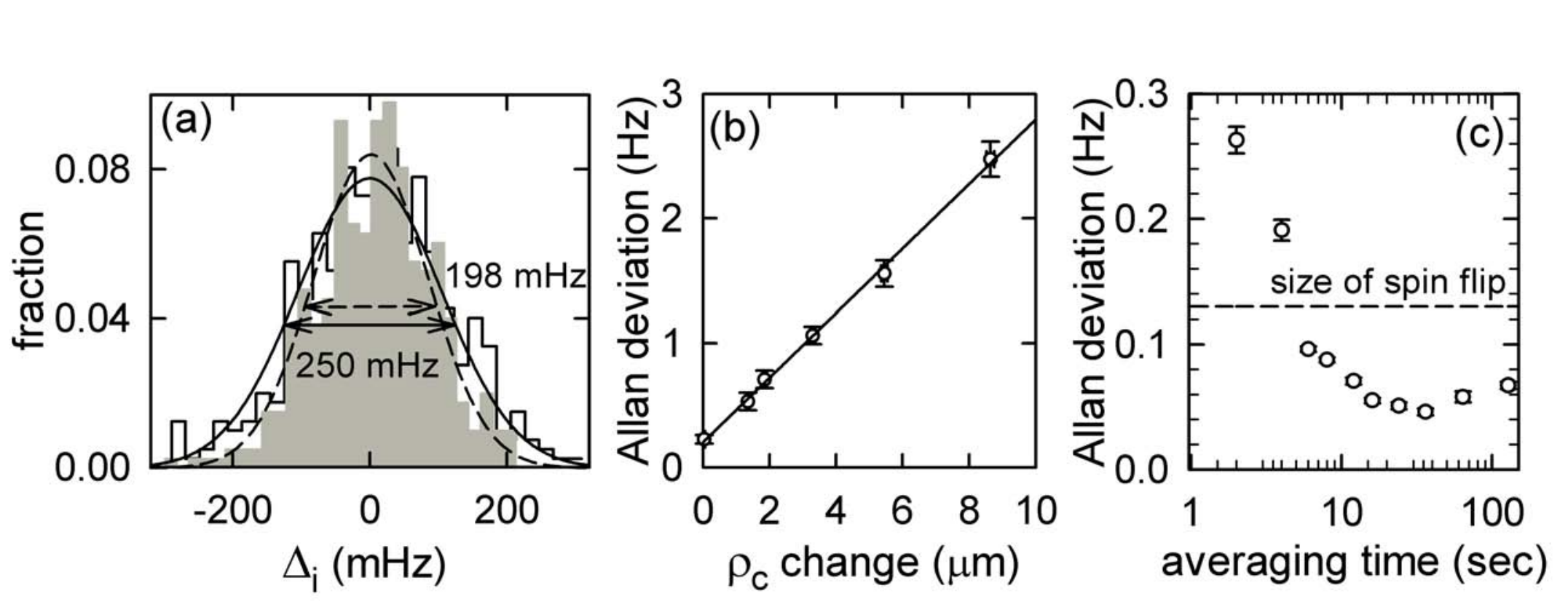}
\caption{(a) Histogram of $\Delta_i$ for a non-resonant (gray) and resonant (black outline) spin flip drive.  Half-widths of 198 mHz and 250 mHz correspond to Allan deviations of $\sigma_0 = 60$ an d $\sigma_f=75$ mHz. (b) Increase in Allan deviation with cyclotron radius. (c) Allan deviation as a function of averaging time for a small cyclotron radius.}
\label{fig:HistogramRadiusAllan}
\end{figure}
}

\newcommand{\MeasurementSequenceAndPowerShiftFigure}{
\begin{figure}[htbp!]
\includegraphics*[width=\columnwidth]{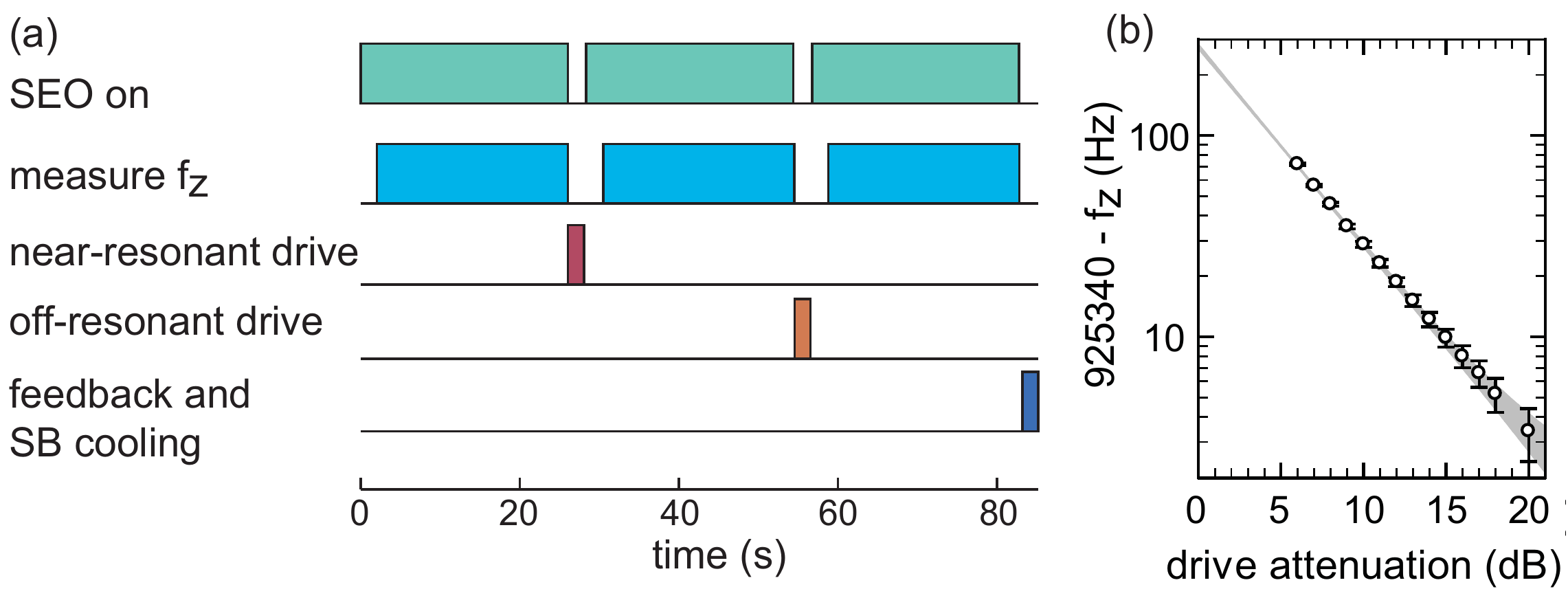}
\caption{(a) Spin measurement sequence. (b) Power shift in $f_z$ due to spin flip drive.}
\label{fig:MeasurementSequence}
\end{figure}
}

\newcommand{\CyclotronAndSpinFlipLineshapesFigure}{
\begin{figure}[htbp!]
\includegraphics*[width=\columnwidth]{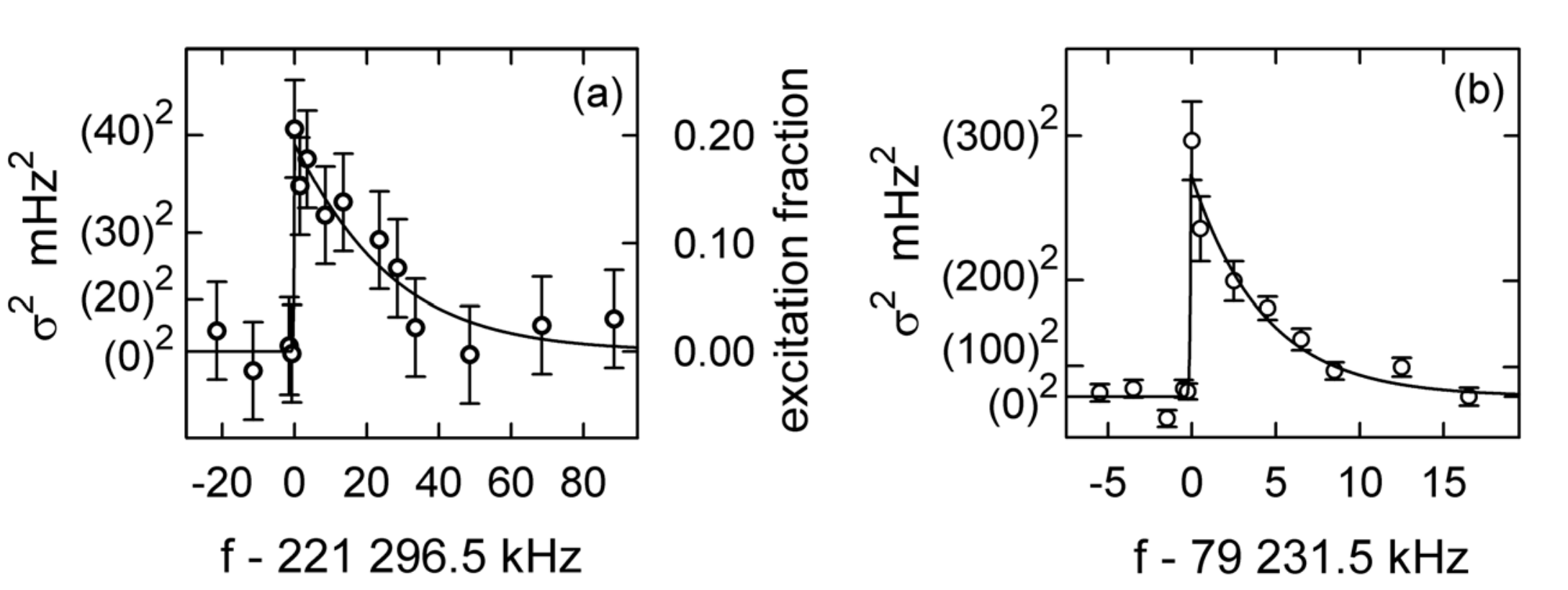}
\caption{(a) The spin line. (b)The cyclotron line.}
\label{fig:CyclotronAndSpinFlipLineshape}
\end{figure}
}

\newcommand{\UncertaintiesTable}{
\begin{table}[htbp!]
    \centering  %\setlength\extrarowheight{2pt}
      \begin{tabular}{llr}

                \hline\hline

                Resonance & ~~~~~Source & ppm\\

                \hline\hline

                spin      & resonance frequency  & 1.7 \\

                \hline

                spin   &  magnetron broadening & 0.7\\
                \hline

                cyclotron~~~~~ & resonance frequency   & 1.6 \\

                \hline

                cyclotron & magnetron broadening & 0.7\\
                \hline \hline

                total & & 2.5 \\

                \hline\hline

            \end{tabular}

        \caption{Significant uncertainties in ppm.}\label{table:Uncertainties}
\end{table}
}

\newcommand{\Harvard}{\affiliation{Dept.\ of Physics, Harvard University, Cambridge, MA 02138}}

\title{Direct Measurement of the Proton Magnetic Moment}

\author{J.\ DiSciacca}
\author{G.\ Gabrielse}\email[Email:]{gabrielse@physics.harvard.edu}
\Harvard

\date{14 Jan. 2012 (submitted to PRL);  31 Jan. 2012 (accepted by PRL).    }

\begin{abstract} %600 characters including spaces (less than 7 manuscript lines)
       The proton magnetic moment in nuclear magnetons is measured to be $\mu_p/\mu_N \equiv g/2 = 2.792\,846 \pm 0.000\,007$, a 2.5 ppm (parts per million) uncertainty.  The direct determination, using a single proton in a Penning trap, demonstrates the first method that should work as well with an antiproton (\pbar) as with a proton (\p).  This opens the way to measuring the \pbar magnetic moment (whose uncertainty has essentially not been reduced for 20 years)  at least $10^3$ times more precisely.
\end{abstract}


\maketitle

The most precisely measured property of an elementary particle is the magnetic moment  of an electron, $\mu_e$, deduced  to  $3 \times 10^{-13}$  from the quantum jump spectroscopy of the lowest quantum states of a single trapped electron \cite{HarvardMagneticMoment2008}.  The moment  was measured in Bohr magnetons, $\mu_e/\mu_B \equiv - g_e/2$ (with $\mu_B=e\hbar/2m_e$ for electron charge $-e$ and mass $m_e$).  The measurement method works for positrons and electrons.  Efforts are thus underway to measure the positron moment (now known 15 times less precisely \cite{DehmeltMagneticMoment}) at the electron precision --  to test lepton CPT invariance to an unprecedented precision.

Applying such one particle methods to measuring the magnetic moment of a \pbar or \p (with mass $m_p$) is challenging.  Nuclear moments scale naturally as the smaller nuclear magneton $\mu_N,$  with $\mu_N/\mu_B = m_e/m_p \sim 1/2000$.
Three measurements and two theoretical corrections \cite{MITProtonMoment,CODATA2006} together determine
$\mu_p/\mu_N$ to 0.01 ppm (Fig.~\ref{fig:MagneticMomentHistory}), but  one of the measurements  relies upon a  hydrogen maser so this method cannot be used for  \pbar .  No $\mu_p$ measurement uses a  method applicable to both \pbar and \p. For more than 20 years the \pbar magnetic moment has been deduced from exotic atom structure, to a precision that has remained at only 3000 ppm \cite{1988PbarMoment,2009PbarMoment} (Fig.~\ref{fig:MagneticMomentHistory}).

\MagneticMomentHistoryFigure

This Letter demonstrates a one-particle method equally applicable to \pbar and    \p, opening the way to a baryon CPT test made by directly comparing  \pbar and p magnetic moments.   The proton  moment in nuclear magnetons is the ratio of its spin and cyclotron frequencies,
\begin{equation}
\frac{\mu_p}{\mu_N} \equiv \frac{g_p}{2} = \frac{f_s}{f_c}.
\label{eq:ProtonMagneticMoment}
\end{equation}
Our measurements of these frequencies with a single trapped \p determine $\mu_p/\mu_N$ to 2.5 ppm, with a value consistent with the indirect determination \cite{MITProtonMoment,CODATA2006}. The possibility to use a single trapped \pbar for precise measurements was established when $f_c$ for a trapped \pbar was measured to $10^{-10}$ \cite{FinalPbarMass}.   It now seems possible to apply the measurement method demonstrated here to determine the \pbar magnetic moment $10^3$ times more precisely than do exotic atom measurements \cite{1988PbarMoment,2009PbarMoment}.  An additional precision increase of $10^3$ or more should be possible once a single \pbar or \p spin flip is resolved.

\PbarMeasurementTrapFigure

For the measurement of $\mu_p$, a single \p is suspended  at the center of the cylindrically symmetric ``analysis trap'' (Fig.~\ref{fig:PbarMeasurementTrap}a).  Its stacked ring electrodes are made of OFE copper or iron, with a 3 mm inner diameter and an evaporated gold layer.  The electrodes and surrounding vacuum container are cooled to 4.2 K by a thermal connection to liquid helium.  Cryopumping of the closed system made the vacuum better than $5 \times 10^{-17}$ Torr in a similar system \cite{PbarMass}, so collisions are not important here. Appropriate potentials applied to electrodes with a carefully chosen relative geometry \cite{OpenTrap} make a very good electrostatic quadrupole near the trap center with open access to the trap interior from either end.

In a vertical magnetic field $\mathbf{B} \approx 5 \,\zhat$ Tesla, a trapped proton's circular cyclotron motion is perpendicular to {\bf B} with a frequency $f_+ = 79.232$ MHz
slightly shifted from $f_c$  by the electrostatic potential.  The proton also oscillates parallel to {\bf B} at about $f_z =
\,925$ kHz.    The proton's third motion is a circular magnetron motion, also perpendicular to
{\bf B}, at the much lower frequency $f_- = 5.395$ kHz.  The spin precession frequency is $f_s=221.30$ MHz.

Driving forces  flip the spin and make cyclotron transitions.  Spin flips require a magnetic field perpendicular to {\bf B} that oscillates at approximately $f_s$.  This field is generated by  currents sent through halves of a compensation electrode (Fig.~\ref{fig:PbarMeasurementTrap}c).   Cyclotron transitions require an electric field perpendicular to {\bf B} that oscillates at approximately $f_+$.  This field is generated by potentials applied across halves of a compensation electrode (Fig.~\ref{fig:PbarMeasurementTrap}d).

Shifts in $f_z$ reveal changes in the cyclotron, spin and magnetron quantum numbers $n$, $m_s$ and $\ell$ \cite{Review},
\begin{equation} \frac{\Delta f_z}{f_z} \approx \frac{\hbar
 \beta_2}{4 \pi m_p |B| f_- } \left( n + \frac{1}{2} + \frac{g_p m_s}{2} + \frac{f_-}{f_+} (\ell + \frac{1}{2})
\right).  \label{eq:FrequencyShift}
\end{equation}
The shifts (50 mHz per cyclotron quanta and 130 mHz for a spin flip) arise when a magnetic bottle gradient,
\begin{equation}
\Delta {\bf B} = \beta_2 [(z^2-\rho^2/2){\bf \hat{z}} - z\rho \boldsymbol{\hat{\rho}}],
\end{equation}
from a saturated iron ring (Fig.~\ref{fig:PbarMeasurementTrap}a) interacts with cyclotron, magnetron and spin moments $\mu \zhat$.  The effective $f_z$ shifts because the electrostatic axial oscillator Hamiltonian going as $f_z^2 z^2$ acquires an additional term  going as $\mu z^2$.  The bottle strength, $\beta_2 = 2.9 \times 10^5$ T/m$^2$, is 190 times that used to detect electron spin flips \cite{HarvardMagneticMoment2008}  to compensate for the small size of the nuclear moments.

A proton is initially loaded into a coaxial trap just above the analysis trap of Fig.~\ref{fig:PbarMeasurementTrap}. Its cyclotron motion induces currents in and comes to thermal equilibrium with a cold damping circuit attached to the trap.  The \p is then transferred to the analysis trap by adjusting electrode potentials to make an axial potential well that moves adiabatically down into the analysis trap.

Two methods are used to measure  the $\Delta f_z$ of Eq.~\ref{eq:FrequencyShift} in the analysis trap, though the choice of which method to use in which context is more historical than necessary at the current precision.  The first (used to detect cyclotron transitions with the weakest possible driving force) takes $\Delta f_z$ to be the shift of the frequency at which noise in a detection circuit is canceled by the signal from the proton axial motion that it drives \cite{DehmeltWalls1968}.  The second (used to detect spin flips) takes $\Delta f_z$ to be the shift in the frequency of a self-excited oscillator (SEO) \cite{OneProtonSelfExcitedOscillator}.  The SEO oscillation arises when amplified signal from the proton's axial oscillation is fed back to drive the \p into a steady-state oscillation.  The detected $f_z$ is first used to check if the cyclotron radius is below 0.3 $\mu m$, a shift $\Delta f_z < 50$ Hz.  If not, the \p is returned to the precision trap for cyclotron damping as needed to select a low cyclotron energy.

Spin and cyclotron measurements are based on sequences of deviations $\Delta_i \equiv f_{i+1} - f_i$, with $i=1,\cdots,N$.  The $f_i$ are a series of averages of the axial frequency over a chosen averaging time.  A histogram of deviations $\Delta_i$ (e.g.\ Fig.~\ref{fig:HistogramRadiusAllan}a)   is characterized by an Allan variance,
\begin{equation}
\sigma^2 = \sum_{i=1}^{N} \Delta_i^2/(2N)
\label{eq:AllanDeviation}
\end{equation}
(often used to describe the stability of frequency sources).  The Allan deviation $\sigma$  is the square root of the variance.

\HistogramRadiusAllanFigure

The Allan variance is $\sigma_0^2$ when no nearly resonant spin or cyclotron drive is applied (gray histogram in Fig.~\ref{fig:HistogramRadiusAllan}a).
The source of this scatter is not yet well understood, as has been discussed \cite{OneProtonSelfExcitedOscillator}.  When a nearly resonant drive at frequency $f$ induces spin flips or cyclotron transitions, the Allan variance increases slightly to $\sigma_f^2=\sigma_0^2 + \sigma^2$ (outline histogram in Fig.~\ref{fig:HistogramRadiusAllan}a).  The small increase, $\sigma^2$, reveals spin or cyclotron resonance. The measured $\sigma_0$ increases with cyclotron radius (Fig.~\ref{fig:HistogramRadiusAllan}b).  It is minimized by selecting a \p with a cyclotron radius below 0.3 $\mu$m, as described.   The measured Allan deviation is then minimum for an averaging time $\tau \approx 30$ s when the SEO is used (Fig.~\ref{fig:HistogramRadiusAllan}c) and for a longer $\tau$ when the noise shorting method is used.

\MeasurementSequenceAndPowerShiftFigure

The cyclotron and spin resonances are well known to be threshold resonances \cite{BrownLineshape,Review}.  A driving force has no effect below a resonance frequency ($f_+$ or $f_s$ here).  The transition rate between quantum states increases abruptly to its maximum at the resonant frequency. Above this threshold there is a distribution  cyclotron or spin frequencies at which these motion can be driven.  These correspond to the distribution of $B$ sampled by the thermal axial motion of the \p (in thermal equilibrium with the axial detection circuit) within the magnetic bottle gradient.

No natural linewidth broadens the sharp threshold edge because the spin and cyclotron motions are not damped in the analysis trap.  The superconducting solenoid produces a stable B that does not significantly smear the edge.  A small broadening arising because sideband cooling (of
magnetron motion coupled to  axial motion)
 selects different values from a distribution of magnetron radii (explored in detail in \cite{OneProtonSelfExcitedOscillator}) is added as ``magnetron broadening'' uncertainty in Table \ref{table:Uncertainties}.

Fig.~\ref{fig:MeasurementSequence} shows the cycle repeated to look for the spin frequency, $f_s$. With the SEO stabilized for 2 s, the SEO frequency is averaged for 24 seconds to get $f_1$.  With the SEO off, a nearly-resonant spin flip drive at frequency $f$ is applied for 2 s.  After the SEO is back on for 2 s its average frequency $f_2$ is measured.  As a control,  a spin drive detuned 50 kHz from resonance is next applied with the SEO off.  It is detuned rather than off to check for secondary effects of the drive.  After the average $f_3$ is measured, 2 s of sideband cooling and feedback cooling keep the magnetron radius from growing \cite{OneProtonSelfExcitedOscillator}.

The cycle in Fig.~\ref{fig:MeasurementSequence} is repeated for typically 36 hours for each drive frequency in Fig.~\ref{fig:CyclotronAndSpinFlipLineshape}a. The Allan deviation  $\sigma_f$ for the sequence of deviations $\Delta_f=f_2-f_1$ represents the effect of fluctuations when a near-resonant spin drive is applied.  The Allan deviation $\sigma_0$ for the sequence of deviations $\Delta_0=f_3-f_2$ represents fluctuations when no near-resonant drive is applied. The spin lineshape in Fig.~\ref{fig:CyclotronAndSpinFlipLineshape}a shows $\sigma^2 = \sigma_f^2 - \sigma_0^2$ vs.\ drive frequency.

The expected sharp threshold that indicates the spin frequency $f_s$ is clearly visible.  The uncertainty in the resonance frequency in Table \ref{table:Uncertainties} is the half-width of the edge.  Smaller frequency steps that might have produced a more precise measurement were not investigated.  The scale to the right in Fig.~\ref{fig:CyclotronAndSpinFlipLineshape}a is the average probability that the spin drive pulse makes a spin flip.  A similar spin resonance is observed in a competing experiment \cite{MainzSpinFlips}, but the fractional half-width of the whole resonance line is about 20 times wider and the fractional uncertainty specified is about 100 times larger.
\CyclotronAndSpinFlipLineshapesFigure

%\CyclotronAndSpinFlipLineshapesVerticalFigure

Matching a pulsed 221 MHz drive so that the needed current goes through required electrode (Fig.~\ref{fig:PbarMeasurementTrap}c) in a cryogenic vacuum enclosure is challenging.  The strong drive applied (because the matching is not optimized) observably shifts $f_z$ as a function of spin drive power (Fig.~\ref{fig:MeasurementSequence}b), presumably because the average trapping potential is slightly modified. The shift from  the strongest drive in Fig.~\ref{fig:MeasurementSequence}b  is still too small to contribute to the uncertainty Table \ref{table:Uncertainties}, however.   Better matching should  produce the same current with less applied drive and shift.

The basic idea of the cyclotron frequency measurement is much the same as for the spin frequency.
The applied drive is weak enough that in 6 hours it causes no detectable growth in the average cyclotron radius and energy even for a resonant drive. The resonant drive is just strong enough to increase the measured Allan variance, $\sigma_f^2$. The cyclotron lineshape (Fig.~\ref{fig:CyclotronAndSpinFlipLineshape}b) shows clearly the expected sharp threshold at the trap cyclotron frequency, $f_+$. The uncertainty in Table \ref{table:Uncertainties} is the half-width of the edge. Smaller frequency steps that might have produced a more precise measurement were not investigated.

For each of the drive frequencies represented in the cyclotron lineshape in Fig.~\ref{fig:CyclotronAndSpinFlipLineshape}b a cyclotron drive is applied continuously for about 6 hours.  This initial approach was adopted to find the weakest useful cyclotron drive and was continued because it worked well. Deviations $\Delta_i$ between consecutive 80 s $f_z$ averages are plotted as a histogram, and characterized by an Allan variance, $\sigma_f^2$.  The $\sigma_0^2$ subtracted off to get $\sigma^2$ uses measurements below the threshold resonance.

We utilize no fits to expected resonance lineshapes for this measurement.  However, we note that the
spin lineshape fits well to the Brownian motion lineshape \cite{BrownLineshape} expected for magnetic field fluctuations caused by thermal axial motion within a magnetic bottle gradient upon a spin 1/2 system.  An axial temperature of 8 K is extracted from the fit, consistent with measurements using a magnetron method detailed in \cite{OneProtonSelfExcitedOscillator}.  With no expected lineshape yet available for the cyclotron resonance, we note that the cyclotron line  fits well to the expected spin lineshape but with an axial temperature of 4 K.
A proper diffusion treatment of the way that a cyclotron drive moves population between cyclotron states is needed.

The magnetic moment in nuclear magnetons is a ratio of frequencies (Eq.~\ref{eq:ProtonMagneticMoment}).  The free space cyclotron frequency, $f_c = e B/(2 \pi m_p)$, is needed while trap eigenfrequencies $f_+$, $f_z$ and $f_m$ are measured directly.  The Brown-Gabrielse invariance theorem, $f_c^2 = f_+^2 + f_z^2 + f_m^2$ \cite{InvarianceTheorem} determines $f_c$ from the eigenfrequencies of an (unavoidably) imperfect Penning trap.

The directly measured proton magnetic moment is
\begin{equation}
\frac{\mu_p}{\mu_N} = \frac{g}{2} = 2.792\,846 \pm 0.000\,007~~~~~[2.5~\rm{ppm}].
\label{eq:Result}
\end{equation}
Uncertainty sources are summarized in Table~\ref{table:Uncertainties}.  Frequency uncertainties are the half-widths of the sharp edges in the lineshapes, as discussed.  The magnetron linewidth uncertainty comes from the distribution of magnetron radii following sideband cooling, as discussed.  All other known uncertainties are too small to show up in this table.

\UncertaintiesTable

The measurement of $\mu_p/\mu_N$ agrees  within 0.2 standard deviations with a 0.01 ppm determination.  The latter is less direct in that three experiments and two theoretical corrections are required, using
\begin{equation}
\frac{\mu_p}{\mu_N} \equiv \frac{g_p}{2} = \frac{\mu_e}{\mu_B} \frac{m_p}{m_e} \frac{\mu_p(H)}{\mu_e(H)}  \frac{\mu_e(H)}{\mu_e}   \frac{\mu_p}{\mu_p(H)}.
\label{eq:gFactorProton}
\end{equation}
The largest uncertainty (0.01 ppm) is for the measured ratio of bound moments, $\mu_p(H)/\mu_e(H)$, measured with a hydrogen maser \cite{MITProtonMoment} that would be difficult to duplicate with antimatter.  The two theoretical corrections between bound and free moments are known ten times more precisely \cite{CODATA1998} .  Both the mentioned $\mu_e/\mu_B$ \cite{HarvardMagneticMoment2008} and $m_p/m_e$ \cite{ElectronMassFromBoundg}  are measured much more precisely still.

This measurement method is the first that can be applied to  \pbar and \p . The magnetic moments of this particle-antiparticle pair could now be compared $10^3$ times more precisely than previously possible (Fig.~\ref{fig:MagneticMomentHistory}) as a test of CPT invariance with a baryon system.

%\MagneticMomentHistoryFigure

A yet illusive goal of this effort \cite{SelfExcitedOscillator} and another \cite{MainzSpinFlips} is to resolve a single spin flip of a trapped \p . The spin could be flipped in a trap with a very small magnetic gradient and thermal broadening, and transferred down to the analysis trap only to determine the spin state. The Allan deviation realized in Fig.~\ref{fig:HistogramRadiusAllan}c suggests that this goal is not far off. Then $f_s$ could then be determined by quantum jump spectroscopy, as for the electron magnetic moment \cite{HarvardMagneticMoment2008}.   Measuring $f_c$ for a trapped \pbar to better than $10^{-10}$ has already been demonstrated \cite{FinalPbarMass} in a trap with a small magnetic gradient.
A comparison of the \pbar and \p magnetic moments that is improved by a factor of a million or more seems possible.  This would add a second precise CPT test with baryons to the $9 \times 10^{-11}$ comparison of the charge-to-mass ratios of \pbar and p \cite{FinalPbarMass}.

In conclusion, a direct measurement of the proton magnetic moment to 2.5  ppm is made with a single \p suspended in a Penning trap.  The measurement is consistent with a more precise determination that uses several experimental and theoretical inputs.  The measurement method is the first that can be applied to a \pbar or \p.  It should now be possible to compare the magnetic moments of \pbar and \p  a thousand times more precisely than has been possible so far, with another thousand-fold improvement factor to be realized when spin flips of a single \p are individually resolved.

 We are grateful for contributions made earlier by N.\ Guise and recently by M.\ Marshall, and for their comments on the manuscript. This work was supported by the AMO programs of the NSF and the AFOSR.

\newcommand{\Desktop}{\bibliography{d:/Jerry/Shared/Synchronize/ggrefs2011}}
\newcommand{\Laptop}{\bibliography{c:/Users/gabrielse/Jerry/Shared/Synchronize/ggrefs2011}}
%\Desktop
%\end{document}

%merlin.mbs apsrev4-1.bst 2010-07-25 4.21a (PWD, AO, DPC) hacked
%Control: key (0)
%Control: author (8) initials jnrlst
%Control: editor formatted (1) identically to author
%Control: production of article title (-1) disabled
%Control: page (0) single
%Control: year (1) truncated
%Control: production of eprint (0) enabled
%

\end{document}